\documentclass[manuscript,screen,nonacm]{acmart}
\AtBeginDocument{%
  \providecommand\BibTeX{{%
    \normalfont B\kern-0.5em{\scshape i\kern-0.25em b}\kern-0.8em\TeX}}}

\setcopyright{acmlicensed}
\copyrightyear{2018}
\acmYear{2018}
\acmDOI{XXXXXXX.XXXXXXX}

\acmConference[Conference acronym 'XX]{Make sure to enter the correct
  conference title from your rights confirmation email}{June 03--05,
  2018}{Woodstock, NY}

\acmBooktitle{Woodstock '18: ACM Symposium on Neural Gaze Detection,
 June 03--05, 2018, Woodstock, NY} 
\acmISBN{978-1-4503-XXXX-X/18/06}

\usepackage{soul}
\usepackage{xspace}

\newcounter{observation}

\definecolor{APA_stats}{RGB}{100, 100, 120}

\usepackage{cleveref}

\begin{document}
\settopmatter{printacmref=false}

\title{Untangling Knots: Leveraging LLM for Error Resolution in Computational Notebooks}

\author{Konstantin Grotov}
\authornote{Authors contributed equally to this research.}
\email{konstantin.grotov@jetbrains.com}
\affiliation{%
  \institution{JetBrains Research}
  \country{Germany}
}

\author{Sergey Titov}
\authornotemark[1]
\email{sergey.titov@jetbrains.com}
\affiliation{
  \institution{JetBrains Research}
  \country{Cyprus}
}

\author{Yaroslav Zharov}
\email{yaroslav.zharov@jetbrains.com}
\affiliation{%
  \institution{JetBrains Research}
  \country{Germany}
}

\author{Timofey Bryksin}
\email{timofey.bryksin@jetbrains.com}
\affiliation{%
  \institution{JetBrains Research}
  \country{Cyprus}
}

\renewcommand{\shortauthors}{Grotov and Titov, et al.}
\newcommand{\ct}{~\hl{[]}\xspace}

\begin{abstract}
Computational notebooks became indispensable tools for research-related development, offering unprecedented interactivity and flexibility in the development process. 
However, these benefits come at the cost of reproducibility and an increased potential for bugs.
There are many tools for bug fixing; however, they are generally targeted at the classical linear code.
With the rise of code-fluent Large Language Models, a new stream of smart bug-fixing tools has emerged.
However, the applicability of those tools is still problematic for non-linear computational notebooks.
In this paper, we propose a potential solution for resolving errors in computational notebooks via an iterative LLM-based agent.
We discuss the questions raised by this approach and share a novel dataset of computational notebooks containing bugs to facilitate the research of the proposed approach.

\end{abstract}



\maketitle

\section{Introduction}
Computational notebooks have become a popular medium for development during the last decade, 
especially for data analysis, machine learning~\cite{perkel2018jupyter}, and creating educational~\cite{barba2019teaching}, or scientific content~\cite{perkel2021ten}. One of the main features of computational notebooks is the possibility of a non-linear development process enabled by creating and executing cells with source code in any order. While this feature allows one to efficiently go through hypotheses and experiment a lot~\cite{rule2018exploration}, it causes high code entanglement~\cite{ramasamy2023workflow, rule2018exploration} and, thus, a higher amount of errors in code. As a result, notebooks are struggling with low reproducibility rates~\cite{pimentel2021understanding, pimentel2019large, wang2021restoring}, when notebooks hardly show the same results after re-running or even can not be executed without errors. And on the other hand, the notebooks' code often contains many stylistic errors~\cite{grotov2022large}. 

In the previous works, several approaches were suggested to solve these issues. To mitigate the code entanglement problem, an algorithm was proposed for the preventive linearization~\cite{macke2020fine}. While existing linters improve code quality, they are not adapted for notebooks and usually produce false positive errors. Hence, in the recent studies, additional notebook-specific linters and guidelines~\cite {samuel2020machine} and tools to match them were presented. Also, special versioning systems~\cite{deo2022runtime} were proposed to solve the reproducibility problem. The emergence of coding Large Language Models~\cite{OpenAI_GPT4_2023, roziere2023code, li2023textbooks, jiang2024mixtral} (LLMs) opens new prospects in this direction since these models can dynamically analyze the context, and address errors in a more nuanced way, compared to static analysis tools.

LLMs demonstrated decent bug-fixing abilities in common coding settings --- repositories with regular scripts code in separate files~\cite{10.1145/3611643.3613892}. However, there is not as much work on applying these techniques to the format of notebooks~\cite{mcnutt2023design}. We suggest that solving the bug-fixing problem in the context of notebooks will benefit both fields --- LLMs and computational notebooks. For the former, notebooks are an example of an interactive environment like works by~\citet{fan2022minedojo,goss2023polycraft} to test their planning and execution abilities. In turn, LLMs could solve the most crucial problems for notebooks: code quality and reproducibility.

In this work, we take the first steps toward such a solution. 
Namely, we 
\begin{enumerate}
    \item collect and share with the community a dataset~\cite{JupyterErrorsDataset} of Python projects that contain computational notebooks with at least one thrown exception;
    \item outline the agent-based~\cite{shavit2023practices, durante2024agent} approach for LLM usage in the notebooks and discuss the research questions we deem important for the practical applicability of this approach, as well as the way to estimate the quality of the solution.
\end{enumerate}

\section{Dataset of Notebook Errors}

Since the rise of the popularity of LLMs, they have already been used as error-solving tools~\cite{xia2023automated, zhang2023self}. However, they were not applied to computational notebooks. To start filling that gap, we are sharing the dataset of computational notebooks containing errors~\cite{JupyterErrorsDataset}. The dataset contains 10,000 notebooks collected by taking a subset from a bigger unpublished corpus. The full corpus consists of all repositories with 10 or more stars on GitHub and all repositories containing Kotlin files. Before further filtering, we excluded forks and no longer actively maintained (archived) repositories to keep the variety of code in the dataset; we also removed the repositories with the last commit date before 2020 to have as many of the modern errors as possible and since popular datasets from \citet{grotov2022large} and \citet{pimentel2019large} provide a good subset of older notebooks. To gather the notebooks, we took the repositories with Python or Jupyter marked as primary languages, collected all notebooks containing errors, and randomly took the first 10,000 notebooks out of 12,609 from 4,636 repositories. We classified a notebook as containing errors if any of the cells' outputs resulted in a Python exception. The detection of these errors was facilitated by using the \texttt{output\_type} property found within the JSON source of the notebook cells. The GitHub snapshot was accurate as of February 21, 2024. All the repositories in the dataset have a permissive license for future research (MIT, Apache-2.0, BSD-3-Clause)~\cite{golubev2020study}.

\begin{figure}[h]
    \centering
\includegraphics[width=\linewidth]{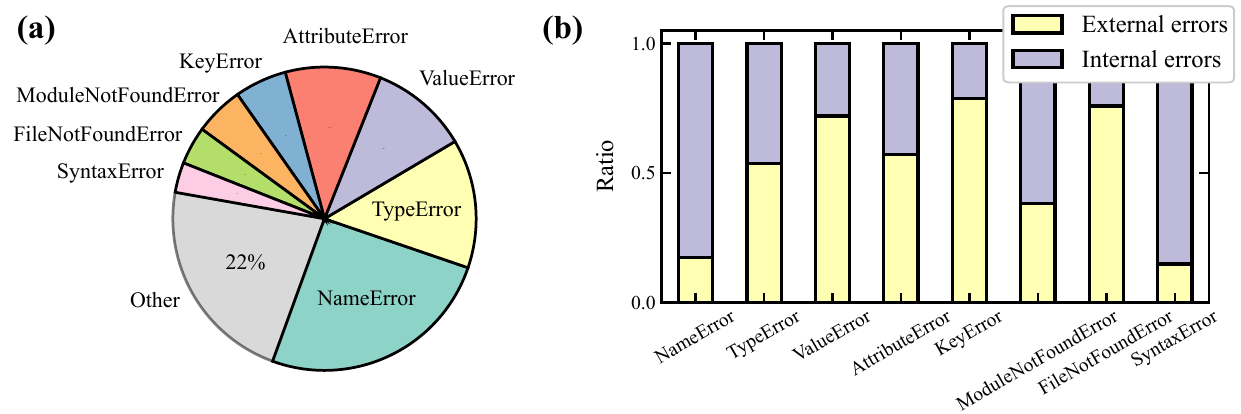}
    \caption{(a) Distribution of top-8 most common errors in GitHub notebooks. (b) The ratio of internal and external errors for every of the top-8 error types.}
    \label{fig:error-distribution}
\end{figure}

To that end, we analyzed the distribution of the most common error types in the collected dataset. We selected all cells that contain errors in the output. Then, for every error, we determined the location of the code that raised the exception; we denote an error as internal if this location is within the notebook and as external otherwise. We determined every error's type using the default Python exception notation. Even though the dataset presents various error types (354 different types), most of the errors belong to a few classes. As illustrated in~\Cref{fig:error-distribution} (a), the distribution of diverse error types aligns with previous studies on notebook reproducibility~\cite{pimentel2019large}. Notably, most errors (78\% of all errors found in the dataset) fall into eight categories. It is worth mentioning that the rest of the categories are sparse, and 282 of them contain less than 5 errors each. However, some of these are pretty narrow and understandable, such as \texttt{SyntaxError} or \texttt{ModuleNotFoundError}, while the top four error types are pretty general and could appear in various circumstances. These top-4 errors can be linked to diverse low-level issues commonly encountered in the non-linear development process~\cite{pimentel2021understanding} used in notebooks, as well as the loss of focus associated with it. For instance, a \texttt{NameError} could occur when the execution order becomes non-linear, and cells in the notebook are used before they are initialized. Furthermore, \texttt{TypeError} and \texttt{ValueError} errors might arise due to incorrect modifications of objects when the execution reverts back to a linear order. For instance, these errors could arise from the use of temporary variables that have the same name and are reassigned to different parts of the notebook. Also, our visual analysis of randomly chosen errors shows that many errors result from incorrect usage of packages and methods or their incompatibility with the installed environment. It's also important to mention that, as shown in~\Cref{fig:error-distribution} (b), most of the mentioned errors are external (caused by the incorrect behavior outside of the notebook, for example, in an imported package), so the gathering additional context is valuable for solving these errors. To facilitate further research, we have made the dataset publicly available and shared the analysis code on Hugging Face~\cite{JupyterErrorsDataset}.

\section{Future Work}

In this section, we briefly outline how to solve such a problem using the approach of a developing field of AI agents. Following this, we pose the research questions that we consider vital for the success of this approach.

\subsection{Error Solving using LLM Agents} 
The field of LLM-based AI agents gained traction recently. Such agents have shown abilities to engage with software engineering tasks~\cite{tufano2024autodev,si2024design2code}, interact with web environments~\cite{drouin2024workarena,zhou2023webarena}, or operate an embodied agent~\cite{wang2023describe}. We suggest employing such agents for iterative error resolution in computational notebooks. The core concept is to allow the agent to code and execute cells, utilizing the notebook's natural interactivity. The agent, therefore, will be able to expand and refine its context before solving the error.

Expanding the context by utilizing methods akin to ones that are used for regular code~\cite{wang2023intervenor}, such as incorporating notebook feedback on the proposed resolution or drawing from data sources and the notebook itself, can indeed be insightful. However, the interactive nature of notebooks might offer additional value. 

The AI agent we envision gains control only after an error occurs and the user seeks a solution. This approach allows the agent to investigate the environment independently and will not destructively interfere with the developers' workflow. As mentioned, this boundary-free exploration is achievable by creating and executing temporary cells and processing the feedback. This provides an avenue for yielding supplemental contextual information which cannot be explicitly present.

\subsection{Research Questions} We pose several research questions for ourselves and the community to solve. 
We think solving these questions will streamline the further development of LLM-based AI agents for the computational notebooks' bugfixing.

\emph{\textbf{RQ1:} How to develop a secure playground?} The security aspect is vital as the environment should provide interaction and error-solving from the agent side, as well as reversibility to the initial user's codebase. We are considering various ways to implement this, from duplicating a notebook and running it in a sandbox to railguard the agent's actions. 

\emph{\textbf{RQ2:} Which metrics should be used to specifically evaluate an agent's performance in a computational notebook environment?} While we can certainly apply existing metrics to make comparisons among agents within a notebook, these measures may not always provide a meaningful view of performance given the context of a notebook environment. More specifically, even if a cell shows no errors, it is possible that the outcome may not align with the developer's original goal due to contextual differences.

\emph{\textbf{RQ3:} What tools can (or should) an agent utilize in a computational notebook?} The expansion of the toolset increases the length of the context window occupied by the tool description. Since agents are, naturally, iterative, the context window can quickly be overflown by the dialogue. This leads to a desire to find a minimal yet complete set of tools. While we mentioned several possible tools for extending the context behind the exception message, we hypothesize that it is beneficial to study the impact of different tools on the agent performance.

\emph{\textbf{RQ4:} Is it possible to accurately solve errors in computational notebooks using Open-Source LLMs?} Open models have already shown decent results for correcting errors in code~\cite{xia2023automated}, but their performance on computational notebooks has not yet been studied. Transitioning to open models could allow one to solve errors without the need for code sharing with third parties. Moreover, it will allow fine-tuning according to the relevant context.

\emph{\textbf{RQ5:} Is the interaction between agents useful for error solving in computational notebooks?} One of the questions open to discussion is whether AI agents perform better as standalone error solvers or when interacting with other agents. Due to the potential for productive communication among agents possessing diverse specific features, we posit that these agents may yield more precise solutions~\cite{iqbal2019actor}. Additionally, interaction with the user may be helpful~\cite{guo2021abgcoqa}.

\section{Conclusion}

In this paper, we considered the task of resolving errors in computational notebooks.
We discussed the possibility of solving such issues with the interactive LLM-based agents.
To facilitate further research, we stated questions that we consider essential for future developments.
Alongside this, we collected, published, and analyzed a dataset of notebooks containing errors.
We hypothesize that solving errors in computational notebooks with agents can be beneficial for research both on AI agents and human-notebook interaction.

\bibliographystyle{ACM-Reference-Format}
\bibliography{bibliography}

\end{document}